\documentclass[preprint,12pt]{elsarticle}

\usepackage{amsthm} % theorems AMS style
\usepackage{amssymb} % math symbols
\usepackage{amstext} % typeset text in math environments
\usepackage{amsmath} % AMS math­e­mat­i­cal fa­cil­i­ties
\usepackage{bm} % bold symbols in math
\usepackage{graphicx} % enhanced support for graphics
\usepackage{epstopdf} % con­vert EPS to 'en­cap­su­lated' PDF us­ing ghostscript
\usepackage{subcaption} % support for subcaptions
\usepackage{float}
\usepackage{tikz}
\usepackage{lineno,hyperref}
\usepackage[capitalize]{cleveref} % in­tel­li­gent cross-ref­er­enc­ing
\usepackage[makeroom]{cancel} %cancel terms
\usepackage{multirow} 
\usepackage{booktabs} 
\usepackage{lscape}
\usepackage[normalem]{ulem}
\usepackage{xcolor,soul}
\usepackage[framemethod=tikz]{mdframed}
\usepackage{enumitem}

\modulolinenumbers[5]

\crefalias{subequation}{equation} % subequations counter = equations counter
\crefalias{eqnarray}{equation} % subequations counter = equations counter
\crefformat{pluraleq}{Eqs.~(#2#1#3)} % defining "plural" equations  
\Crefformat{pluraleq}{Equations~(#2#1#3)} % defining "plural" equations  

\def\bal#1\nal{\begin{align}#1\end{align}}
\def\bala#1\nala{\begin{align*}#1\end{align*}}
\def\bsub#1\nsub{\begin{subequations}#1\end{subequations}}
\newcommand{\f}{\frac}
\newcommand{\ux}{{\bm x}}
\newcommand{\un}{{\bm n}}
\newcommand{\unab}{{\bf \nabla}}

\newcommand{\uom}{{\bf \Omega}}

\journal{arXiv}

 \biboptions{sort&compress}
\begin{document}

\begin{frontmatter}

\title{Transport Synthetic Acceleration for the Solution of the One-Speed \\ Nonclassical Spectral S$_N$ Equations in Slab Geometry
}

\author[osu]{J.K.~Patel\fnref{patel}}
\author[uerj]{L.R.C.~Moraes\fnref{moraes}}
\author[osu]{R.~Vasques\corref{cor1}}
\author[uerj]{R.C.~Barros\fnref{barros}}

\address[osu]{The Ohio State University, Department of Mechanical and Aerospace Engineering\\ 201 W. 19th Avenue, Columbus, OH 43210}
%\address[ufrgs]{Universidade Federal do Rio Grande do Sul, Departamento de Matem\'atica Pura e Aplicada -- IME \\ Av.~Bento Gon\c calves 9500, 91509-900, Porto Alegre, RS, Brazil.}
\address[uerj]{Universidade do Estado do Rio de Janeiro, Departamento de Modelagem Computacional -- IPRJ\\ Rua Bonfim 25, 28625-570, Nova Friburgo, RJ, Brazil}

\cortext[cor1]{Corresponding author: richard.vasques@fulbrightmail.org; Tel: (510) 340 0930\\
Postal address: The Ohio State University, Department of Mechanical and Aerospace Engineering, 201 W. 19th Avenue, Columbus, OH 43210}
\fntext[patel]{patel.3545@osu.edu}
\fntext[moraes]{leonardrcmoraes@gmail.com}
\fntext[barros]{ricardob@iprj.uerj.br}

\begin{abstract}

The nonclassical transport equation models particle transport processes in which the particle flux does not decrease as an exponential function of the particle's free-path.
Recently, a spectral approach was developed to generate nonclassical spectral S$_N$ equations, which can be numerically solved in a deterministic fashion using classical numerical techniques.
This paper introduces a transport synthetic acceleration procedure to speed up the iteration scheme for the solution of the monoenergetic slab-geometry nonclassical spectral S$_N$ equations.
We present numerical results that confirm the benefit of the acceleration procedure for this class of problems.
 
\end{abstract}


\end{frontmatter}

\setcounter{section}{0}
\setcounter{equation}{0} 

\section{Introduction}\label{sec1}
\setcounter{equation}{0} 

In the \textit{classical} theory of linear particle transport, the incremental probability that a particle at position $\ux = (x,y,z)$ will experience a collision while traveling an incremental distance $ds$ in the background material is given by
\bal\label{1.1}
dp = \sigma_t(\ux)ds\,,
\nal
where $\sigma_t$ represents the macroscopic total cross section \cite{case}.
The implicit assumption is that $\sigma_t$ is independent of the particle's direction-of-flight $\uom$ and of the particle's free-path $s$, defined as the distance traveled by the particle since its previous interaction (birth or scattering).
This assumption leads to the particle flux being exponentially attenuated (Beer-Lambert law). 
We remark that extending the results discussed here to include energy- or frequency-dependence is straightforward.

The theory of \textit{nonclassical} particle transport employs a generalized form of the linear Boltzmann equation to model processes in which the particle flux is \textit{not} attenuated exponentially.
This area has been significantly researched in recent years
\cite{lar_07,larvas_11,fragou_10,vaslar_14a,deo_14,siam_15,aml_16,frasun_16,vassla_17,cam_17,larfra_17,jctt_18,deon_18a,deon_18b}.
Originally introduced to describe photon transport in the Earth's cloudy atmosphere  \cite{davmar_10,kryber_13,davxu_14,xudav_16,davxu_18,davxu_18b}, it has found its way to other applications, including nuclear engineering \cite{vaslar_09,vas_13,vaslar_14b,vassla_16,vaskry_17}
and computer graphics
\cite{wren_17,jarabo,bitterli}.
Furthermore, an analogous theory yielding a similar kinetic equation has been independently derived for the periodic Lorentz gas
by Marklof and Str\"ombergsson \cite{marstr_10,marstr_11,marstr_14,marstr_15}
and by Golse (cf.~\cite{gol_12}).

The nonclassical transport equation allows the \textit{nonclassical} macroscopic total cross section $\Sigma_t$ to be a function of the particle's free-path, and is defined in an extended phase space that includes $s$ as an independent variable.
If we define
\bal\label{1.2}
P(\ux,\uom,s)ds =\left(
\begin{array}{l}
\text{the probability that a particle released at position $\ux$ in the}\\
\text{direction $\uom$ will experience its next collision while traveling}\\
\text{an incremental interval between $s$ and $s+ds$}
\end{array}
\right)\,,
\nal
then we can define the ensemble average 
\bal\label{1.3}
p(s) = \left< P(\ux,\uom,s) \right>_{(\ux,\uom,\mathcal{R})}
\nal
over all ``release positions" $\ux$ in a realization of the system, all directions $\uom$, and all possible realizations $\mathcal{R}$.
In this case, $p(s)$ represents the free-path distribution function, and the nonclassical cross section $\Sigma_t(s)$ satisfies 
\bal\label{1.4}
p(s) = \Sigma_t(s) e^{-\int_0^s\Sigma_t(s')ds'}\,.
\nal
It is possible to extend this definition to include angular-dependent free-path distributions and cross sections \cite{vaslar_14a}, but in this paper we will restrict ourselves to the case given by \cref{1.4}.
 
 The steady-state, one-speed nonclassical transport equation with isotropic scattering can be written as \cite{larvas_11}
\bsub\label[pluraleq]{1.5}
\bal
&\frac{\partial}{\partial s} \Psi(\ux, \uom, s) + \uom\cdot\unab\Psi(\ux, \uom, s) + \Sigma_t(s) \Psi(\ux, \uom, s) = \label{1.5a}\\
&\qquad\qquad\delta(s)\left[\frac{c}{4\pi} \int_{4\pi} \int_0^{\infty} \Sigma_t(s') \Psi(\ux, \uom', s')ds' d\Omega' + \frac{Q(\ux)}{4\pi}\right]\,,
\hspace{5pt} \ux\in V,\,\, \uom \in 4\pi,\,\,0<s, \nonumber
\nal
where $\Psi$ is the nonclassical angular flux, $c$ is the scattering ratio, and $Q$ is an isotropic internal source.
The Dirac delta function $\delta(s)$ on the right-hand side of \cref{1.5a} represents the fact that a particle that has just undergone scattering or been born will have its free-path value (distance since previous interaction) set to $s =0$.
If we consider vacuum boundaries, \cref{1.5a} is subject to the boundary condition
\bal
\Psi(\ux,\uom, s) = 0\,, \quad \ux \in \partial V,\,\, \un \cdot \uom <0,\,\, 0<s\,.
\nal
\nsub

We remark that, if $\Sigma_t(s)$ is assumed to be independent of $s$, then $\Sigma_t(s) = \sigma_t$ and the free-path distribution in \cref{1.4} reduces to the exponential
\bal\label{1.6}
p(s) = \sigma_t e^{-\sigma_t s}\,.
\nal
In this case, \cref{1.5a} can be shown to reduce to the corresponding classical linear Boltzmann equation 
 \bsub\label[pluraleq]{1.7}
\bal
\uom\cdot\unab\Psi_c(\ux, \uom) + \sigma_t \Psi_c(\ux, \uom) = \frac{c}{4\pi} \int_{4\pi} \sigma_t\Psi_c(\ux, \uom')d\Omega' + \frac{Q(\ux)}{4\pi}\,,
\hspace{10pt} \ux\in V,\,\, \uom \in 4\pi\, ,
\nal
with vacuum boundary condition given by
\bal
\Psi_c(\ux,\uom) = 0\,, \quad \ux \in \partial V,\,\, \un \cdot \uom <0\,.
\nal
\nsub
Here, the classical angular flux $\Psi_c$ is given by
\bal\label{1.8}
\Psi_c(\ux,\uom) = \int_0^\infty \Psi(\ux,\uom,s)ds\,.
\nal

Recently, a spectral approach was developed to represent the nonclassical flux as a series of Laguerre polynomials in the variable $s$ \cite{jcp_20}.
The resulting equation has the form of a classical transport equation that can be solved in a deterministic fashion using traditional methods.
Specifically, the nonclassical solution was obtained using the conventional discrete ordinates (S$_N$) formulation \cite{lewis} and a source iteration (SI) scheme \cite{adalar_02}. 
However, for highly scattering systems the spectral radius of the transport problem can get arbitrarily close to unity \cite{lewis}, and numerical acceleration becomes important. 

The goal of this paper is to introduce transport synthetic acceleration techniques--namely, S$_2$ synthetic acceleration (S$_2$SA)--to speed up the solution of the nonclassical spectral S$_N$ equations.
We also present numerical results that confirm the benefit of using this approach; to our knowledge, this is the first time such acceleration methods are applied to this class of nonclassical spectral problems.

The remainder of the paper is organized as follows.
In \cref{sec2}, we present the nonclassical spectral S$_N$ equations for slab geometry. 
We discuss transport synthetic acceleration in \cref{sec3} and present an iterative method to efficiently solve the nonclassical problem. 
Numerical results are given in \cref{sec4} for problems with both exponential (\cref{sec4.1}) and nonexponential (\cref{sec4.2}) choices of $p(s)$. 
We conclude with a brief discussion in \cref{sec5}. 

\section{Nonclassical Spectral S$_N$ Equations in Slab Geometry}\label{sec2}
\setcounter{equation}{0}

In this section we briefly sketch out the derivation of the one-speed nonclassical spectral S$_N$ equations in slab geometry.
For a detailed derivation, we direct the reader to the work presented in \cite{jcp_20}.

In slab geometry, \cref{1.5} can be written as 
\bsub\label[pluraleq]{2.1}
\bal
&\frac{\partial}{\partial s} \Psi(x, \mu, s) + \mu \frac{\partial}{\partial x} \Psi(x, \mu, s) + \Sigma_t(s) \Psi(x, \mu, s) = \label{2.1a}\\
&\qquad\delta(s)\left[\frac{c}{2} \int_{-1}^1 \int_0^{\infty}  \Sigma_t(s') \Psi(x, \mu', s')ds' d\mu' + \frac{Q(x)}{2}\right]\,,
\hspace{5pt} 0<x<X,\,\, -1<\mu<1,\,\,0<s, \nonumber\\
&\Psi(0, \mu, s) = 0\,, \quad 0<\mu\leq1\,, 0< s\,,\\
&\Psi(X, \mu, s) = 0\,, \quad -1\leq\mu<0\,, 0<s\,,
\nal
\nsub
where $\mu$ is the cosine of the scattering angle. 
\Cref{2.1a} can be written in an equivalent ``initial value" form:
\bsub\label[pluraleq]{2.2}
\bal
&\frac{\partial}{\partial s} \Psi(x, \mu, s) + \mu \frac{\partial}{\partial x} \Psi(x, \mu, s) + \Sigma_t(s) \Psi(x, \mu, s) = 0\,,\\
&\Psi(x,\mu,0) = \frac{c}{2} \int_{-1}^1 \int_0^{\infty}  \Sigma_t(s') \Psi(x, \mu', s')ds' d\mu' + \frac{Q(x)}{2}\,.\label{2.2b}
\nal
\nsub
Note that, due to scattering and internal source being isotropic, the right-hand side of \cref{2.2b} does not depend on $\mu$.

Defining $\psi$ such that 
\bal\label{2.3}
\Psi(x, \mu, s) \equiv \psi(x,\mu,s)e^{-\int_0^s  \Sigma_t(s')ds'}\,,
\nal
we can rewrite the nonclassical problem as
\bsub\label[pluraleq]{2.4}
\bal
  &\frac{\partial}{\partial s}\psi(x, \mu, s) + \mu \frac{\partial}{\partial x} \psi(x, \mu, s) = 0,\label{2.4a}\\
  & \psi(x, \mu, 0) = \frac{c}{2} \int_{-1}^1  \int_0^{\infty}  p(s')\psi(x, \mu', s')ds'd\mu' + \frac{Q(x)}{2}\,,
\nal
where $p(s)$ is given by \cref{1.4}.
This problem has the vacuum boundary conditions
\bal
&\psi(0, \mu, s) = 0\,, \quad 0<\mu\leq1\,, 0< s\,,\\
&\psi(X, \mu, s) = 0\,, \quad -1\leq\mu<0\,, 0<s\,.
\nal
\nsub
Next, we write $\psi$ as a truncated series of Laguerre polynomials in $s$:
\bal\label{2.5}
\psi(x, \mu, s) = \sum_{m=0}^M \psi_m(x, \mu) L_m(s),
\nal
where $L_m(s)$ is the Laguerre polynomial of order $m$ and $M$ is the expansion (truncation) order. 
The Laguerre polynomials $\{ L_m(s)\}$ are orthogonal with respect to the weight function $e^{-s}$ and satisfy $\f{d}{ds}L_m(s) = \left(\f{d}{ds}-1\right)L_{m-1}(s)$ for $m>0$ \cite{hoc_72}.
We introduce this expansion in the nonclassical problem and perform the following steps \cite{jcp_20}: (i) multiply \cref{2.4a} by $e^{-s}L_m(s)$; (ii) integrate from $0$ to $\infty$ with respect to $s$; and (iii) use the properties of the Laguerre polynomials to simplify the result.
This procedure returns the following nonclassical spectral problem:
\bsub
\bal
  &\mu \frac{\partial}{\partial x} \psi_m(x, \mu) + \psi_m(x, \mu) = S(x) + \frac{Q(x)}{2} - \sum_{j=0}^{m-1} \psi_j(x, \mu), \quad m = 0, 1, ..., M\,,\\
  &\psi_m(0, \mu) = 0\,, \quad 0<\mu\leq1\,,m = 0, 1, ..., M\,,\\
&\psi_m(X, \mu) = 0\,, \quad -1\leq\mu<0\,,m = 0, 1, ..., M\,,
\nal
where the in-scattering term $S(x)$ (the scattering source) is given by
\bal
S(x) = \frac{c}{2} \int_{-1}^1  \sum_{k=0}^M \psi_k(x, \mu')\left[\int_0^{\infty}p(s)  L_k(s)ds\right] d\mu'\,.
\nal
\nsub
The nonclassical angular flux $\Psi$ is recovered from \cref{2.3,2.5}.
The classical angular flux $\Psi_c$ is obtained using \cref{1.8}, such that
\bal
\Psi_c(x,\mu) = \int_0^\infty \Psi(x,\mu,s) ds = \sum_{m=0}^M \psi_m(x,\mu)\int_0^\infty L_m(s)e^{-\int_0^s\Sigma_t(s')ds'}ds\,. 
\nal

Finally, using the discrete ordinates formulation \cite{lewis}, we can write the nonclassical spectral S$_N$ equations
\bsub\label[pluraleq]{2.8}
\bal
  &\mu_n \frac{d}{d x} \psi_{m,n}(x) + \psi_{m,n}(x) = S(x) + \frac{Q(x)}{2} - \sum_{j=0}^{m-1} \psi_{j,n}(x), \label{2.8a} \\
&  \hspace{250pt} m = 0, 1, ..., M,\, n = 1, 2, ..., N\,,\nonumber\\
  &\psi_{m,n}(0) = 0\,, \quad m = 0, 1, ..., M,\,n = 1, 2, ..., \f{N}{2}\,,\\
&\psi_{m,n}(X) = 0\,, \quad m = 0, 1, ..., M,\,n = \f{N}{2}+1, ..., N\,,\\
& S(x) = \frac{c}{2} \sum_{n=1}^N \omega_n  \sum_{k=0}^M \psi_{k,n}(x)\left[\int_0^{\infty}p(s) L_k(s)ds\right]\,,\\
&\Psi_{c_n}(x) = \sum_{m=0}^M \psi_{m,n}(x)\int_0^\infty L_m(s)e^{-\int_0^s\Sigma_t(s')ds'}ds\,, \quad n = 1, 2, ..., N\,. \label{2.8e} 
\nal
\nsub
Here, the cosine of the scattering angle $\mu$ has been discretized in $N$ discrete values $\mu_n$.
Thus, $\psi_{m,n}(x) = \psi_{m}(x,\mu_n)$, $\Psi_{c_n}(x) = \Psi_c(x,\mu_n)$, and the angular integral has been approximated by the angular quadrature formula with weights $\omega_n$.

\section{Source Iteration and Synthetic Acceleration}\label{sec3}
\setcounter{equation}{0} 

To solve the nonclassical spectral S$_N$ equations using standard source iteration \cite{adalar_02}, we lag the scattering source on the right-hand side of \cref{2.8a}: 
\bsub
\bal
  \mu_n \frac{d}{d x} \psi_{m,n}^{i+1}(x) + \psi_{m,n}^{i+1}(x) = S^i(x) + \frac{Q(x)}{2} - \sum_{j=0}^{m-1} \psi_{j, n}^{i+1}(x),
\nal
where $i$ is the iteration index and
\bal
 S^i(x) = \frac{c}{2} \sum_{n=1}^N \omega_n   \sum_{k=0}^M \psi_{k,n}^i(x) \left[\int_0^{\infty}p(s) L_k(s)ds\right]\,.
\nal
\nsub
In order to accelerate the convergence of this approach, the iterative scheme is broken into multiple stages.

Standard synthetic acceleration methods consist of two stages.
The first stage is a single transport sweep.
The second stage is error-correction, which uses an approximation of the error equation to estimate the error at each iteration.
Our synthetic acceleration scheme has the following steps:
\begin{itemize}
\item[1.] \textit{Determine the new ``half iterate" $\psi^{i+\f{1}{2}}$ (solution estimate) using one transport sweep.}

This is done by solving
\bal\label{3.2}
  \mu_n \frac{d}{d x} \psi_{m,n}^{i+\f{1}{2}}(x) + \psi_{m,n}^{i+\f{1}{2}}(x) = S^i(x) + \frac{Q(x)}{2} - \sum_{j=0}^{m-1} \psi_{j, n}^{i+\f{1}{2}}(x).
\nal
\item[2.] \textit{Approximate the error $\epsilon^{i+1}$ in this half iterate using an approximation to the error equation (error estimate).}

To do that, we first subtract \cref{3.2} from the exact equation \cref{2.8a}, then add and subtract $\psi_{k,n}^{i+\f{1}{2}}(x)$ to the in-scattering term on the right-hand side.
This yields
\bsub\label[pluraleq]{3.3}
\bal
  \mu_n \frac{d}{d x} \left(\psi_{m,n}(x)-\psi_{m,n}^{i+\f{1}{2}}(x)\right) &+ \left(\psi_{m,n}(x)-\psi_{m,n}^{i+\f{1}{2}}(x)\right) = \\
&\qquad\qquad \left(S(x)-S^i(x)\right) - \sum_{j=0}^{m-1} \left(\psi_{j,n}(x)-\psi_{j,n}^{i+\f{1}{2}}(x)\right)\,, \nonumber
\nal
where
\bal
 S(x)-S^{i}(x) = \frac{c}{2} \sum_{n=1}^N \omega_n   \sum_{k=0}^M &\left(\psi_{k,n}(x)-\psi_{k,n}^{i+\f{1}{2}}(x)+ \right.\\
 &\qquad\qquad\left.\psi_{k,n}^{i+\f{1}{2}}(x)-\psi_{k,n}^i(x)\right)\left[\int_0^{\infty}p(s) L_k(s)ds\right]\,.\nonumber
\nal
\nsub
Defining the error $\epsilon^{i+1}_{m,n}$ as
\bal
\epsilon^{i+1}_{m,n}(x) \equiv \psi_{m,n}(x) - \psi_{m,n}^{i+\f{1}{2}}(x)\,,
\nal
we rewrite \cref{3.3} as
\bsub\label[pluraleq]{3.5}
\bal
  \mu_n \frac{d}{d x} \epsilon^{i+1}_{m,n}(x) + \epsilon^{i+1}_{m,n}(x) - S^{{i+1},\epsilon}(x) =\label{3.5a} \left(S^{i+\f{1}{2}}(x)-S^i(x)\right)-  \sum_{j=0}^{m-1}\epsilon^{i+1}_{j,n}(x)\,, 
\nal
with
\bal
 S^{{i+1},\epsilon}(x)= \frac{c}{2} \sum_{n=1}^N \omega_n \sum_{k=0}^M \epsilon^{i+1}_{k,n}(x) \left[\int_0^{\infty}p(s)L_k(s)ds\right]\,.
\nal
\nsub
We solve \cref{3.5} and obtain the error estimate $\epsilon^{i+1}$.
\item[3.] \textit{Correct the solution estimate using the error estimate.}

The corrected solution estimate $\psi^{i+1}$ is given by
\bal
\psi^{i+1}_{m,n}(x) = \psi^{i+\f{1}{2}}_{m,n}(x) + \epsilon^{i+1}_{m,n}(x)\,.
\nal
\item[4.] \textit{Check for convergence and loop back if necessary.}
\end{itemize}

\noindent We remark that this transport synthetic acceleration procedure accelerates each one of the $M$ Laguerre moments of the angular flux.
In this paper, we have chosen to approximate the error estimate in \cref{3.5} by setting $N=2$, thus applying S$_2$ synthetic acceleration (S$_2$SA). 

\section{Numerical Results}\label{sec4}
\setcounter{equation}{0} 

In this section we provide numerical results that confirm the benefit of using transport synthetic acceleration for the iterative numerical solution of the nonclassical spectral S$_N$ equations (\ref{2.8}).
For validation purposes, we first apply this nonclassical approach to solve a transport problem with an exponential $p(s)$, which leads to classical transport.
Then, we proceed to solve a nonclassical transport problem that mimics diffusion, with a nonexponential $p(s)$.

For all numerical experiments in this section we use the Gauss-Legendre angular quadrature \cite{burden} with $N=16$ for \cref{2.8} and $N=2$ for \cref{3.5}, thus solving the nonclassical spectral S$_{16}$ equations using S$_2$ synthetic acceleration.
We discretize the spatial variable into 200 elements and use the linear discontinuous Galerkin finite element method \cite{adams}.
Furthermore, the improper integrals $\int_0^\infty (\cdot)ds$ in these equations are calculated numerically in the same fashion as in \cite{jcp_20}: the upper limit is truncated to $1.5$ times the length of the slab, and a Gauss-Legendre quadrature is used to solve them.
Here, we set the order of this quadrature to $M$, the same order as the Laguerre expansion.  

The stopping criterion adopted is that the relative deviations between two consecutive estimates of the classical scalar flux
\bal\label{4.1}
\Phi(x) = \sum_{n=1}^N\omega_n\Psi_{c_n}(x)
\nal
in each point of the spatial discretization grid need to be smaller than or equal to a prescribed positive constant $\xi$.
For all our calculations we fix $\xi=10^{-6}$, such that the stopping criterion is given by
\bal
\frac{||\Phi^{i+1}(x) - \Phi^{i}(x)||}{||\Phi^{i}(x)||} \leq \xi\,.
\nal

\subsection{Exponential $p(s)$}
\label{sec4.1}

To validate the approach, we use the nonclassical method to solve a transport problem in which $p(s)$ is given by the exponential function provided in \cref{1.6}. 
This yields \cite{larvas_11}
\bal\label{4.3}
\Sigma_t(s) = \f{p(s)}{\int_s^\infty p(s')ds'} = \f{\sigma_t e^{-\sigma_t s}}{\int_s^\infty \sigma_t e^{-\sigma_t s'}ds'} =  \sigma_t \text{ (independent of $s$).} 
\nal  
In this case, the flux $\Psi_{c_n}$ given by \cref{2.8e} should match the one obtained by solving the corresponding \textit{classical} S$_N$ transport problem
\bsub\label[pluraleq]{4.4}
\bal
&\mu_n\f{d}{dx}\Psi_{c_n}(x) + \sigma_t\Psi_{c_n}(x) = \f{c}{2}\sigma_t\sum_{n=1}^N\omega_n\Psi_{c_n}(x) + \f{Q(x)}{2}\,,\quad 0<x<X\,, n=1,2,...,N\,,\\
&\Psi_{c_n}(0) = 0\,, \quad n=1,2,...,\f{N}{2}\,,\\
&\Psi_{c_n}(X) = 0\,, \quad n=\f{N}{2}+1,...,N\,.
\nal
\nsub

\begin{figure}[!t]
\centering
  \includegraphics[scale=0.1]{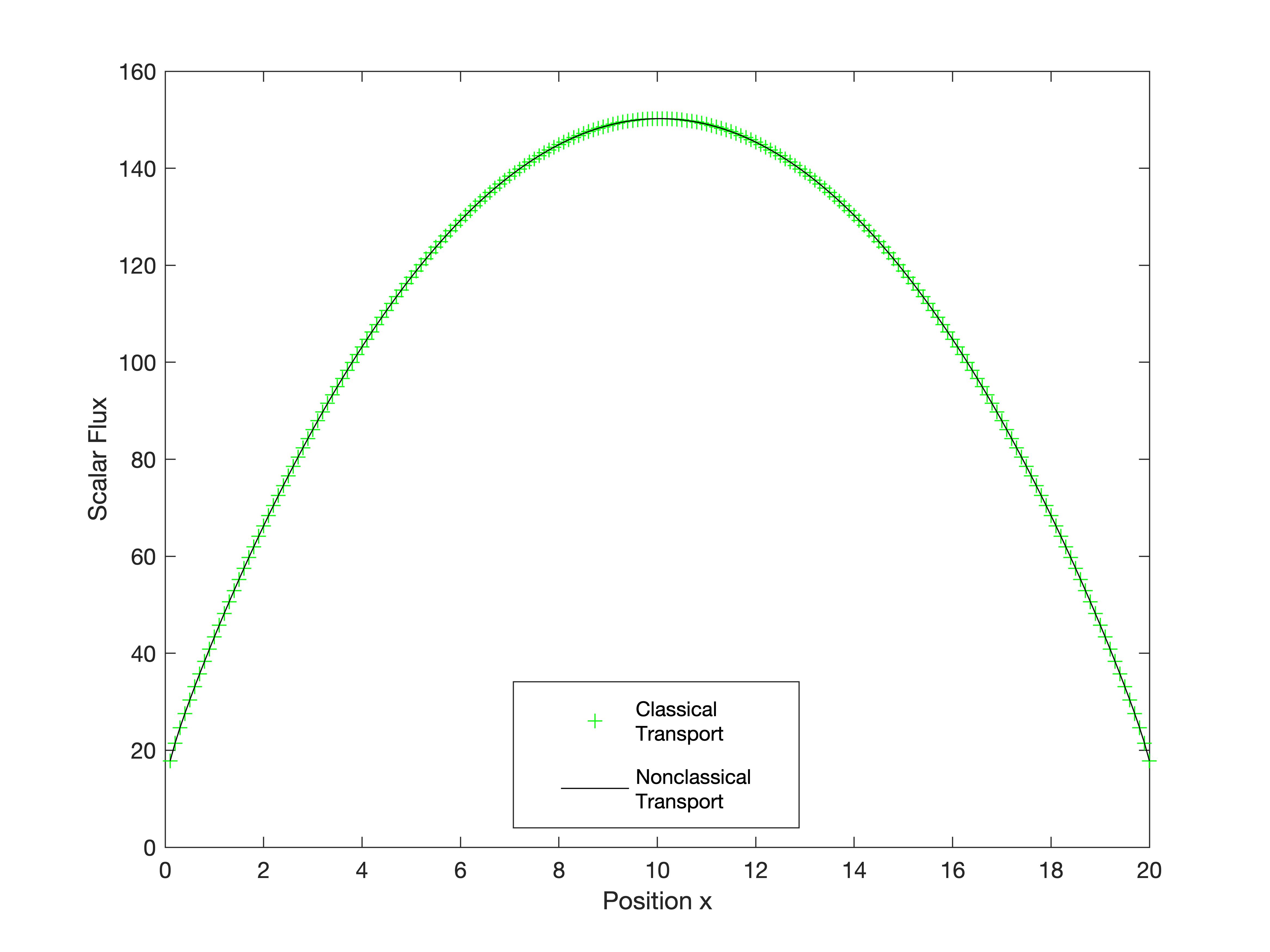}
   \caption{Scalar flux generated by the solution of classical and nonclassical transport equations}
  \label{fig1}
\end{figure}

Let us consider a slab of length $X=20$, total cross section $\sigma_t = 1.0$, scattering ratio $c=0.999$, and internal source $Q(x) = 1.0$, and let us assume
the truncation order of the Laguerre expansion to be $M=10$. 
\Cref{fig1} depicts the scalar flux obtained when solving the nonclassical (\ref{2.8}) and classical (\ref{4.4}) problems.
As expected, the solutions match each other. 

\begin{table}[!b]
\begin{center}
\caption{Convergence Data for Nonclassical Transport with Exponential $p(s)$}\label{tab1}
\begin{tabular}{||c||c|c||c|c||}
\hline
\hline
 & \multicolumn{2}{|c||}{\textbf{Number of}} & \multicolumn{2}{c||}{\textbf{Spectral}}\\
 & \multicolumn{2}{|c||}{\textbf{Iterations}} & \multicolumn{2}{c||}{\textbf{Radius}} \\
 \cline{2-5} 
\vspace{-10pt} &&&&\\
 $\mathbf{c}$ & \textbf{SI} & \textbf{S$_2$SA} & \textbf{SI} & \textbf{S$_2$SA}  \\
 \hline
 \vspace{-10pt} &&&&\\
 0.8 & 56 & 6 & 0.7997 & 0.1328 \\
 0.9 & 110 & 6 & 0.8997 & 0.1565 \\
 0.99 & 906 & 6 & 0.9899 & 0.1748 \\
 0.999 & 6439 & 6 & 0.9989 & 0.1685\\
 \hline\hline
\end{tabular}
\end{center}
\end{table}

Next, we compare the iteration count and spectral radius for stand-alone source iteration (SI) and transport synthetic acceleration (S$_2$SA) for the nonclassical method.
Once again, we set $Q(x)=1.0$, and $p(s)$ and $\Sigma_t(s)$ are given respectively by Eqs.~(\ref{1.6}) and (\ref{4.3}), with $\sigma_t =1.0$.
However, this time we increase the domain size to $X= 200$.
We assume the truncation order of the Laguerre expansion to be $M=50$, and vary the scattering ratio $c$ from 0.8 to 0.999.
\Cref{tab1} presents the number of iterations and the spectral radius for each case.
As expected, we observe a significant reduction in the spectral radius and iteration count, with the number of iterations decreasing 3 orders of magnitude for the highest scattering ratio of $c=0.999$.

\subsection{Nonexponential $p(s)$}
\label{sec4.2}

Let us consider the diffusion equation in a homogeneous slab
\bsub\label[pluraleq]{4.5}
\bal
-\f{1}{3\sigma_t}\f{d^2}{dx^2} \phi(x) + (1-c)\sigma_t \phi(x) = Q(x)\,,
\nal
with Marshak boundary conditions \cite{bell}
\bal
&\phi(0)-\f{2}{3\sigma_t}\f{d}{dx}\phi(0) = 0\,,\\
&\phi(X)+\f{2}{3\sigma_t}\f{d}{dx}\phi(X) = 0\,.
\nal
\nsub
Here, $\phi$ is the (diffusion) scalar flux.

If the free-path distribution $p(s)$ is given by the nonexponential function
\bal\label{4.6}
p(s) = 3\sigma_t^2 s\, e^{-\sqrt{3}\sigma_t s},
\nal
it has been shown that the collision-rate density $\sigma_t\phi(x)$ of the diffusion problem, given by \cref{4.5}, will match the \textit{nonclassical} collision-rate density \cite{siam_15,aml_16} 
\bal
f(x) = \int_0^\infty\Sigma_t(s)\int_{-1}^1\Psi(x,\mu,s)d\mu ds,
\nal
where $\Psi(x,\mu,s)$ is the solution of the nonclassical problem given by \cref{2.1}, and
\bal\label{4.8}
\Sigma_t(s) = \f{p(s)}{\int_s^\infty p(s')ds'} = \f{3\sigma_t^2 s e^{-\sqrt{3}\sigma_t s}}{\int_s^\infty 3\sigma_t^2 s' e^{-\sqrt{3}\sigma_t s'} ds'} =   \f{3\sigma_t^2 s}{1+\sqrt{3}\sigma_t s}\,.
\nal  

\begin{figure}[!t]
\centering
  \includegraphics[scale=0.1]{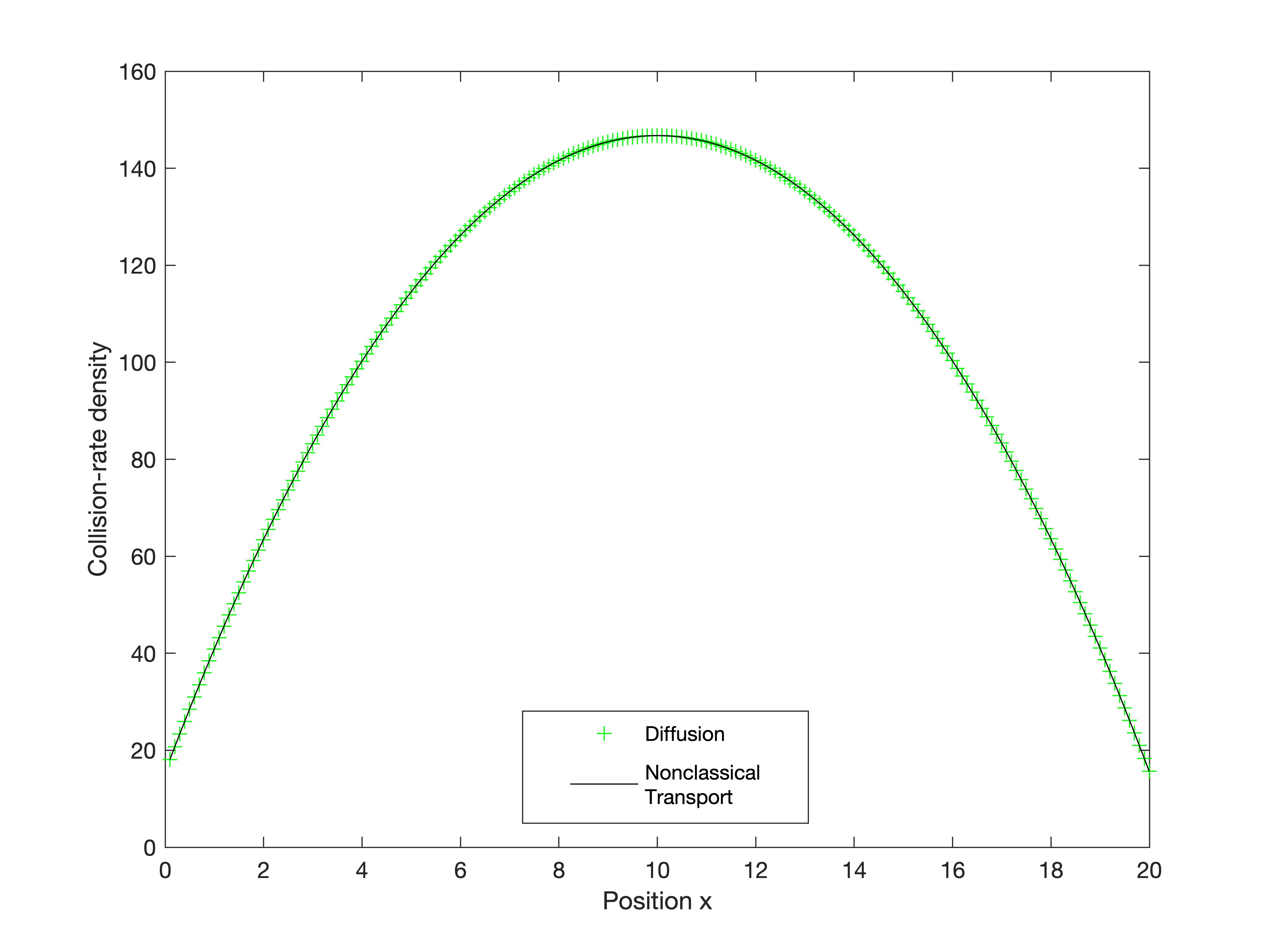}
   \caption{Collision-rate density generated with diffusion and nonclassical approaches}
  \label{fig2}
\end{figure}

Once again, let us consider a slab of length $X=20$, $\sigma_t = 1.0$, $c=0.999$, and $Q(x) = 1.0$.
\Cref{fig2} shows a comparison between the collision-rate densities of the diffusion problem (\cref{4.5}) and the nonclassical spectral S$_N$ method (\cref{2.8}), with the latter being given by
\bal
f(x) = \sum_{n=1}^N \omega_n \sum_{m=0}^{M} \psi_{m,n}(x)\int_0^\infty p(s)L_m(s)ds,
\nal 
where $M=10$.
As in the previous case, the solutions match as expected.

\begin{table}[!b]
\begin{center}
\caption{Convergence Data for Nonclassical Transport with Nonexponential $p(s)$}\label{tab2}
\begin{tabular}{||c||c|c||c|c||}
\hline
\hline
 & \multicolumn{2}{|c||}{\textbf{Number of}} & \multicolumn{2}{c||}{\textbf{Spectral}}\\
 & \multicolumn{2}{|c||}{\textbf{Iterations}} & \multicolumn{2}{c||}{\textbf{Radius}} \\
 \cline{2-5} 
\vspace{-10pt} &&&&\\
 $\mathbf{c}$ & \textbf{SI} & \textbf{S$_2$SA} & \textbf{SI} & \textbf{S$_2$SA}  \\
 \hline
 \vspace{-10pt} &&&&\\
 0.8 & 56 & 6 & 0.7997 & 0.1538 \\
 0.9 & 110 & 7 & 0.8997 & 0.1811 \\
 0.99 & 906 & 6 & 0.9989 & 0.1885 \\
 0.999 & 6443 & 6 & 0.9989 & 0.1802\\
 \hline\hline
\end{tabular}
\end{center}
\end{table}

At this point, we compare the iteration count and spectral radius for stand-alone source iteration (SI) and transport synthetic acceleration (S$_2$SA).
We set $Q(x)=1.0$, and $p(s)$ and $\Sigma_t(s)$ are given respectively by \cref{4.6,4.8}, with $\sigma_t =1.0$.
Once more, we increase the domain size to $X= 200$, and assume
the truncation order of the Laguerre expansion to be $M=50$.
\Cref{tab2} presents the number of iterations and the spectral radius for different choices of the scattering ratio $c$.
Similar to the previous case, there is a reduction in both the spectral radius and iteration count, with a decrease of 3 orders of magnitude in the iterations for the highest scattering ratio.

%to solve a problem in which $pthe numerical approach described in the previous section to solve a classicaltransport problem in slab geometry. The slab is composed of a homogeneous material (material 1), with total cross section t1=1cm−1, as depicted in Fig.1.

\section{Discussion}\label{sec5}
\setcounter{equation}{0} 

We have introduced a transport synthetic acceleration procedure that speeds up the source iteration scheme for the solution of the one-speed nonclassical spectral S$_N$ equations in slab geometry.
Specifically, we used S$_2$ synthetic acceleration to solve nonclassical spectral S$_{16}$ equations for problems involving exponential and nonexponential free-path distributions.
The numerical results successfully confirm the advantage of the method; to our knowledge, this is the first time a numerical acceleration approach is used in this class of nonclassical spectral problems.
Moreover, although we assumed for simplicity monoenergetic transport and isotropic scattering, extending the method to include energy-dependence and anisotropic scattering shall not lead to significant additional theoretical difficulties.

When compared to stand-alone SI, S$_2$SA yields a significant reduction in number of iterations (up to three orders of magnitude) and spectral radii. 
The values of the spectral radius for stand-alone SI  remain virtually unchanged for the exponential and nonexponential cases for a fixed value of the scattering ratio $c$.
However, all spectral radii for S$_2$SA are larger in the nonexponential case than in the exponential case for the same value of $c$, increasing from $6.5\%$ (when $c = 0.999$) to $13.6\%$ (when $c=0.8$).

In fact, we do not see spectral radius values that are exactly consistent with those found when applying corresponding techniques to the classical S$_N$ transport equation \cite{adalar_02}.
This can be attributed to the fact that the nonclassical equation contains an altogether different scattering term, which depends on the free-path $s$.
Although a full convergence analysis is beyond the scope of this paper, we shall perform it in a future work in order to investigate this feature.

\section*{Acknowledgements}

J.~K.~Patel and R.~Vasques acknowledge support under award number NRC-HQ-84-15-G-0024 from the Nuclear Regulatory Commission.
This study was financed in part by the Coordena\c{c}\~ao de
Aperfei\c{c}oamento de Pessoal de N\'ivel Superior - Brasil (CAPES) -
Finance Code 001.
L.~R.~C.~Moraes and R.~C.~Barros also would like to express their gratitude to the support of Conselho Nacional de Desenvolvimento Cient\'ifico e Tecnol\'ogico - Brasil (CNPq) and Funda\c{c}\~ao Carlos Chagas Filho de Amparo \`a Pesquisa do Estado do Rio de Janeiro - Brasil (FAPERJ).

%\section*{References}

\bibliography{JCAM_BIB}

\end{document}